\documentclass[11pt,twoside]{article}

\usepackage{asp2006}
\usepackage{epsf}
\usepackage{graphicx}
\usepackage{lscape}
\usepackage{amsbsy}
\usepackage{amssymb}

\begin{document}
\title{Freeware solutions for spectropolarimetric data reduction}

\author{Fr\'ed\'eric Paletou, and Ludovick L\'eger}
\affil{Universit\'e de Toulouse, Observatoire Midi-Pyr\'en\'ees, LATT
  (CNRS/UMR5572), 14 ave. E. Belin, F-31400 Toulouse, France}

\author{Reza Rezaei} \affil{Kipenheuer Institut f\"ur Sonnenphysik,
  Sch\"oneckstr. 6, D-79104 Freiburg, Germany}

\begin{abstract} 
  Most of the solar physicists use very expensive software for data
  reduction and visualization. We present hereafter a reliable
  freeware solution based on the Python language. This is made
  possible by the association of the latter with a small set of
  additional libraries developed in the scientific community. It
  provides then a very powerful and economical alternative to other
  interactive data languages. Although it can also be used for any
  kind of post-processing of data, we demonstrate the capabities of
  such a set of freeware tools using TH\'eMIS observations of the
  second solar spectrum.
\end{abstract}

\section{Introduction}   
The Python language \citep{vanrossum} is increasingly popular in the
scientific and, more specifically, in the astronomical community. For
instance, it has been already adopted by very large communities
such as the Space Telescope Science Institute (STScI) or by the ALMA
project. It is also very strong among the various tools used and
developed in the frame of the Virtual Observatory endeavour (see
{\tt http://www.ivoa.net/} for instance).

In combination with a small number of specific libraries, we explain
briefly here how it can provide a very powerful interactive data
language which can be run on a number of operating system platforms,
such as Linux, Mac OS X or Windows. The set of Python libraries we
shall describe herafter does not present an interest limited to data
reduction since it can also be perfectly used for any kind of
post-processing like, for instance, the one of output from numerical
simulations.

For the spectropolarimetric data reduction we performed here, we used
a combination of tools from the libraries {\tt PyFITS}, {\tt numarray}
(now {\tt NumPY/SciPy}) and {\tt matplotlib}.

Using data collected by G. Molodij and F. Paletou in May 2000 at the
TH\'eMIS solar telescope \citep{fpgm}, we demonstrate the capabilities
of such tools by extracting the so-called 2nd spectrum of the Sr\,{\sc
i} spectral line at 460.7 nm close to the solar limb, as an
illustrative example \citep[e.g.,][]{stenflo97}.

By reprocessing such data with our Python-based tools, we could easily
reproduce the results published elsewhere \citep{jtbmcvfpgm}.

\begin{figure}
  \centering
  \includegraphics[width=14cm,angle=0]{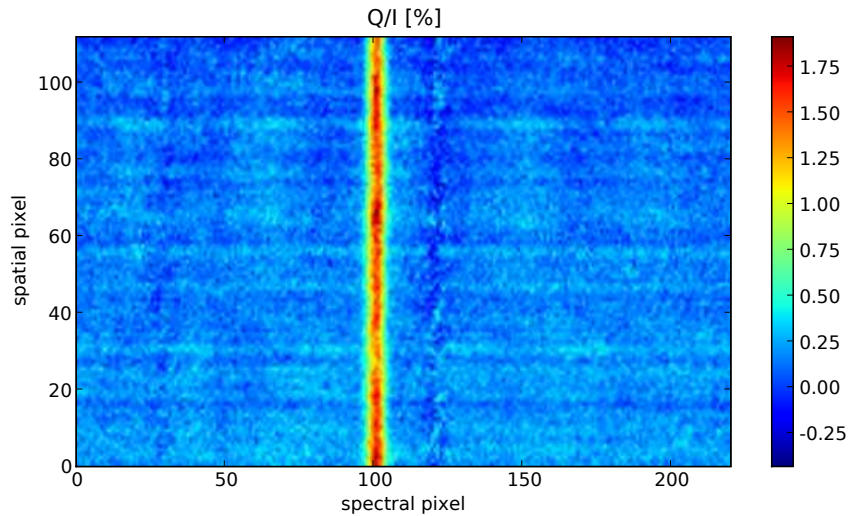}
  \caption{$x-\lambda$ image of the fractional linear polarization
  $Q(\lambda)/I(\lambda)$ in the spectral domain around 460.7 nm.}
\end{figure}

\section{Necessary resources}

In order to deal with FITS format files, the STScI developed -- and
maintains -- the {\tt PyFITS}\footnote{{\tt
http://www.stsci.edu/resources/software\_hardware/pyfits}} library. It
is very easy to install and to use since it is very well
documented. It allows for both the reading of such files, entirely or
by slices, and for the generation of new FITS files; working with
headers is also very easy. However its use implies that another
library such as {\tt numarray} or {\tt NumPy} (see below) for handling
multi-dimensional arrays is already available.

Indeed, for our purpose vector calculations with multi-dimensional
arrays have been performed with the {\tt numarray} library. With the
latter comes along also a number of high-level numerical tools
allowing for linear algebra, statistical analysis, fast fourier
transforms, convolutions or interpolations for instance. However,
since {\tt numarray} will {\em not} be supported anymore after 2007,
we wish to warn the reader to use instead, from now on, the {\tt
NumPy}\footnote{\tt http://numpy.scipy.org/} package for such
scientific calculations \citep[see also][]{oliphant}. An on-line
cookbook and very useful documentation can be found at {\tt
http://www.scipy.org/}.

Finally, for graphical output and figures saving, we used the {\tt
matplotlib}\footnote{\tt http://matplotlib.sourceforge.net/} library
for 2D plots. The default GUI is very convenient, allowing for a
posteriori interactive work on the image, such as area selection and
zooming. Unlike other software, it is also very easy with {\tt
matplotlib} to export images into the most useful formats. The quality
of the output is perfectly suitable for publication \citep[see e.g.,
the figures in][]{leger07}.

The freeware set we adopted was installed and used by us without any
difficulties together with several Linux distributions as well as Mac OS X.

\section{Capabilities and results}

The raw data consisted in a time-sequence of 200 frames taken with the
slit parallel to the solar limb at $\mu \approx 0.1$ while modulating in
time through a sequence of 4 independent polarization states in order
to perform full-Stokes measurements.

Demodulation involving inverse or pseudo-inverse matrix calculation
was very easily coded with the linear algebra package of {\tt NumPy}.
Our older routines (written in IDL) for aligning spectral lines were
also very quickly re-coded using convolution with a finite width kernel
for bisector search, and cubic spline or Fourier interpolations for
shifting the profiles line by line. The adequate functions are in the
{\tt fftpack} and {\tt signal} packages of {\tt SciPy}.

\begin{figure}
  \centering
  \includegraphics[width=12cm,angle=0]{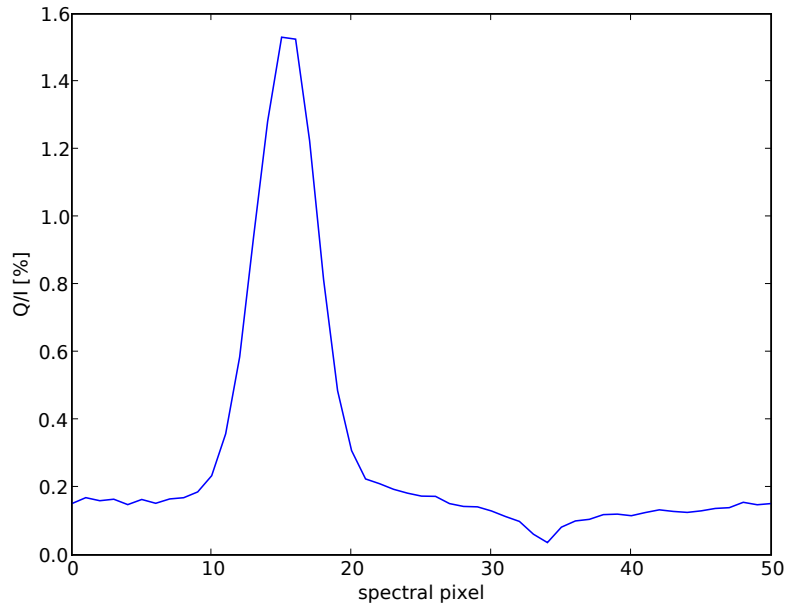}
  \caption{Mean linear polarization profile in the 460.7 nm spectral
  domain. The profile results from the averaging along the slit of the
  map displayed in Fig. 1. The depolarization of a nearby Fe\,{\sc
  i} line at pixel $\approx 33$ can also be noticed.}
\end{figure}

In Fig. 1, we plot the ($x-\lambda$) image corresponding to the
fractional linear polarization $Q(\lambda)/I(\lambda)$ obtained by
recombining the two polarized beams of the MTR@TH\'eMIS \citep{fpgm}
taken at 460.7 nm. The strong peak in the Sr\,{\sc i} spectral line
corresponds to $\approx 1.5\%$. Fig. 2 makes this reading easier since
it is the mean profile obtained by averaging the polarized signal over
the rows of the previous image.

It is in agreement with previous values obtained at very high
polarimetric sensitivity with the ZIMPOL {\sc i} polarimeter attached
to the NSO/Kitt Peak MacMath-Pierce facility \citep{stenflo97}.

As with other data language, these packages permit to build one's own
collection of specific functions and, therefore makes it possible for
any user to constitute its own library. The use of these public and
private resources can be made both from scripts and/or interactively,
using a command line.

One of the advantages of using Python is flexibility. Several other
high-level plotting libraries exist such as DISLIN\footnote{\tt
http://www.mps.mpg.de/dislin/}, or VTK\footnote{\tt
http://www.vtk.org/} for 3D graphics, for instance. And should one be
unsatisfied of some of the functions of {\tt matplotlib}, it would be
very easy to use instead a number of functions from such other libraries.

\section{Conclusions}

The Python language and the numerous scientific and graphic libraries
which are being developed for, already provide very valuable and
powerful tools for data analysis in astrophysics. It is well supported
by increasingly larger communities so, shifting to such freeware tools
appear to us as both reasonable and economical (not to say legal
too...) options yet.

Our experience with the reduction of TH\'eMIS spectropolarimetric data
since 1999 made rather fast, on the timescale of a few weeks only, the
conversion of our former software written in IDL to the
above-mentioned Python-based resources. We could reprocess without any
difficulties whatsoever old data which lead to published results, and
we now reduce our new data with those numerical tools. And finally, we
definitely adopt a proselytizing attitude in favour of them.

\acknowledgements Rafael Manso is highly acknowledged for his
constant interest during this quest for freeware solutions.

\end{document}